\documentclass[usenatbib]{emulateapj}
\usepackage{float}

\shorttitle{The increasing flattening of quiescent galaxies with
  redshift}
\shortauthors{A. R. Hill et al.}
\gdef\new{}
\begin{document}
\title{High redshift massive quiescent Galaxies are as flat as star forming
  galaxies: the flattening of galaxies and the correlation with structural
properties in CANDELS/3D-HST}

\author{Allison R. Hill$^{1}$, Arjen van der Wel$^{2}$, Marijn Franx$^{1}$, Adam Muzzin$^{3}$, Rosalind E. Skelton$^{4}$, Iva Momcheva$^{5}$, Pieter van Dokkum$^{5}$, Katherine E. Whitaker$^{6}$}
\affil{$^{1}$Leiden Observatory, Leiden University, P.O. Box 9513, 2300 RA,
Leiden, The Netherlands}
\affil{$^{2}$Department of Physics and Astronomy, Ghent University, 9000 Gent, Belgium}
\affil{$^{3}$Department of Physics and Astronomy, York University, 4700 Keele St., Toronto, Ontario, Canada, MJ3 1P3}
\affil{$^{4}$South African Astronomical Observatory, PO Box 9, Observatory, Cape Town, 7935, South Africa}
\affil{$^{5}$Astronomy Department, Yale University, New Haven, CT 06511, USA}
\affil{$^{6}$Department of Physics, University of Connecticut, Storrs, CT 06269, USA}

\email{hill@strw.leidenuniv.nl}
\slugcomment{ApJ, in press (accepted Nov 18, 2018)}

\begin{abstract}

We investigate the median flattening of galaxies at
$0.2<z<4.0$ in all five CANDELS/3D-HST fields via the apparent axis
ratio $q$. We separate the sample into bins of redshift, stellar-mass,
s{\'e}rsic index, size, and UVJ determined star-forming state to
discover the most important drivers of the median $q$
($q_{med}$). Quiescent galaxies at $z<1$ and $M_{*}>10^{11}M_{\odot}$
are rounder than those at lower masses, consistent with the hypothesis
that they have grown significantly through dry merging. The massive
quiescent galaxies at higher redshift become flatter, and are as flat
as star forming massive galaxies at $2.5<z<3.5$, consistent with
formation through direct transformations or wet mergers. We find that
in quiescent galaxies, correlations with $q_{med}$ and $M_{*}$, $z$
and $r_{e}$ are driven by the evolution in the s{\'e}rsic index ($n$),
{\new consistent with the growing accumulation of minor mergers at lower
redshift}. Interestingly, $n$ does not drive these trends fully in
star-forming galaxies. Instead, the strongest predictor of $q$ in
star-forming galaxies is the effective radius, where larger galaxies are
flatter. Our findings suggest  that  $q_{med}$ is tracing
bulge-to-total ($B/T$) galaxy ratio which would explain why
smaller/more massive star-forming galaxies are rounder than their
extended/less massive analogues, {\new although it is unclear why
  Sersic index correlates more weakly with flattening for star forming
  galaxies than for quiescent galaxies}

\end{abstract}


\section{Introduction}
\label{sec:intro}

Tracing the morphological evolution of galaxies from photometry is valuable in providing insights into the underlying kinematics of galaxy evolution when time-expensive, high S/N spectra are unavailable. Physical parameters have long been known to broadly couple to Hubble-type \citep[e.g.,][]{roberts1994, blanton2003}, with young, star forming (SF) galaxies exhibiting some form of gas-rich disk or flattened structure and quiescent (Q) galaxies exhibiting older stellar populations in rounder, puffed up ellipticals (although passive disks do make up a small, but not insignificant population of passive galaxies; \citealt[e.g.,][]{bruce2014a}). 

In order to quantify the morphological evolution, various structural parameters have proven to be useful proxies for visual classification. In general, disk galaxies have been associated with a low ($n\sim1$) Sersic index surface brightness profile (or an exponential profile), and elliptical galaxies with a high ($n\sim4$) Sersic index light profile (de Vacouleurs profile). Along with a Sersic parameter, galaxies have also been quantified based on their effective radius, $r_{e}$, and their apparent axis ratio, $q$. 

On a galaxy-by-galaxy basis, $q$ is not in itself a very useful
parameter as it can depend strongly on inclination angle. However,
distributions of $q$ have been used to infer the intrinsic axis ratios
of populations of galaxies separated by their Hubble type
\citep[e.g.,][]{sandage1970, lambas1992} and by mass, star-forming
state and redshift \citep[e.g.,][]{law2012, chang2013,
  vanderwel2014b}. 
For instance, in the local universe,
\citet{lambas1992} found that the elliptical $q$-distribution implied
that these galaxies are intrinsically triaxial as pure oblate/prolate
models could not account for the observed axis ratio distributions.

\citet{vanderwel2014b} and \citet{chang2013} used similar methodology to measure how the distributions evolve with redshift in star-forming and quiescent galaxies. \citet{chang2013} confirmed that the apparent axis ratio distribution of quiescent galaxies at low-$z$ is consistent with intrinsic triaxial shapes, and that this is also true in their high-redshift ($1<z<2.5$) counterparts. They also found that at $z>1$, galaxies with $M_{*}\sim10^{11}M_{\odot}$ exhibited a higher oblate fraction which they interpreted as massive galaxies being comprised of disks in the past, which were destroyed in major-merger events. For lower-mass quiescent galaxies ($M_{*}<10^{10.5}M_{\odot}$), the evolution of the oblate fraction is reversed, with low-mass quiescent galaxies at high-$z$ not having sufficient time to settle into stable disk systems as compared to today. 

In star-forming galaxies, \citet{vanderwel2014b} found that disks are ubiquitous among massive galaxies at all redshifts below $z\sim2$. At lower stellar mass ($M_{*}<10^{10}M_{\odot}$), the fraction of galaxies with elongated intrinsic shapes increases towards higher redshifts and lower masses, and that similar to their low-mass quiescent counter parts discussed in \citet{chang2013}, these galaxies did not have sufficient time to settle into stable disks. This interpretation is supported by kinematic analysis in IFU studies, such as \citet{simons2017a} who find that disordered (i.e. dispersion dominated) motions decreases with decreasing redshift in low-mass star-forming galaxies.

In this study, we choose to investigate the median apparent axis-ratio
($q_{med}$) evolution instead of modelling the distributions and
inferring their intrinsic shapes. We instead, infer the intrinsic
flattening from the median flattening, with the underlying assumption
that the trends in the median encapsulate trends in the larger
population. We caveat this with the fact that many studies who
investigated the apparent axis ratio distribution, $P(q)$, find that a
single morphological type often does not reproduce the observed
$P(q)$, and that the models demand a more heterogeneous population
\citep[e.g.,][]{lambas1992, chang2013, vanderwel2014b}. By using the
$q_{med}$, we can quantify the dependency on other structural
parameters such as $n$, and $r_{e}$ and their evolution. We analyze
how these values change as a function of the star-forming state of
these galaxies and determine what $q_{med}$ is tracing in these
different populations.

{\new We note that the apparent average flattening of a population of
 galaxies is closely related to the average intrinsic flattening defined by
  the ratio of the short axis to the long axis of a galaxy \citep[see,
    e.g.][]{franx1991}. The ratio of intermediate axis to long axis
  influences the apparent flattening only weakly.}

Throughout this article, we assume a $\mathrm{\Lambda}$-CDM cosmology ($H_{0}=\mathrm{70~kms^{-1}Mpc^{-1}}$, $\Omega_{M}=0.3$, and $\Omega_{\Lambda}=0.7$).

\begin{figure*}[ht]
  \begin{center}
  \includegraphics[width=15truecm]{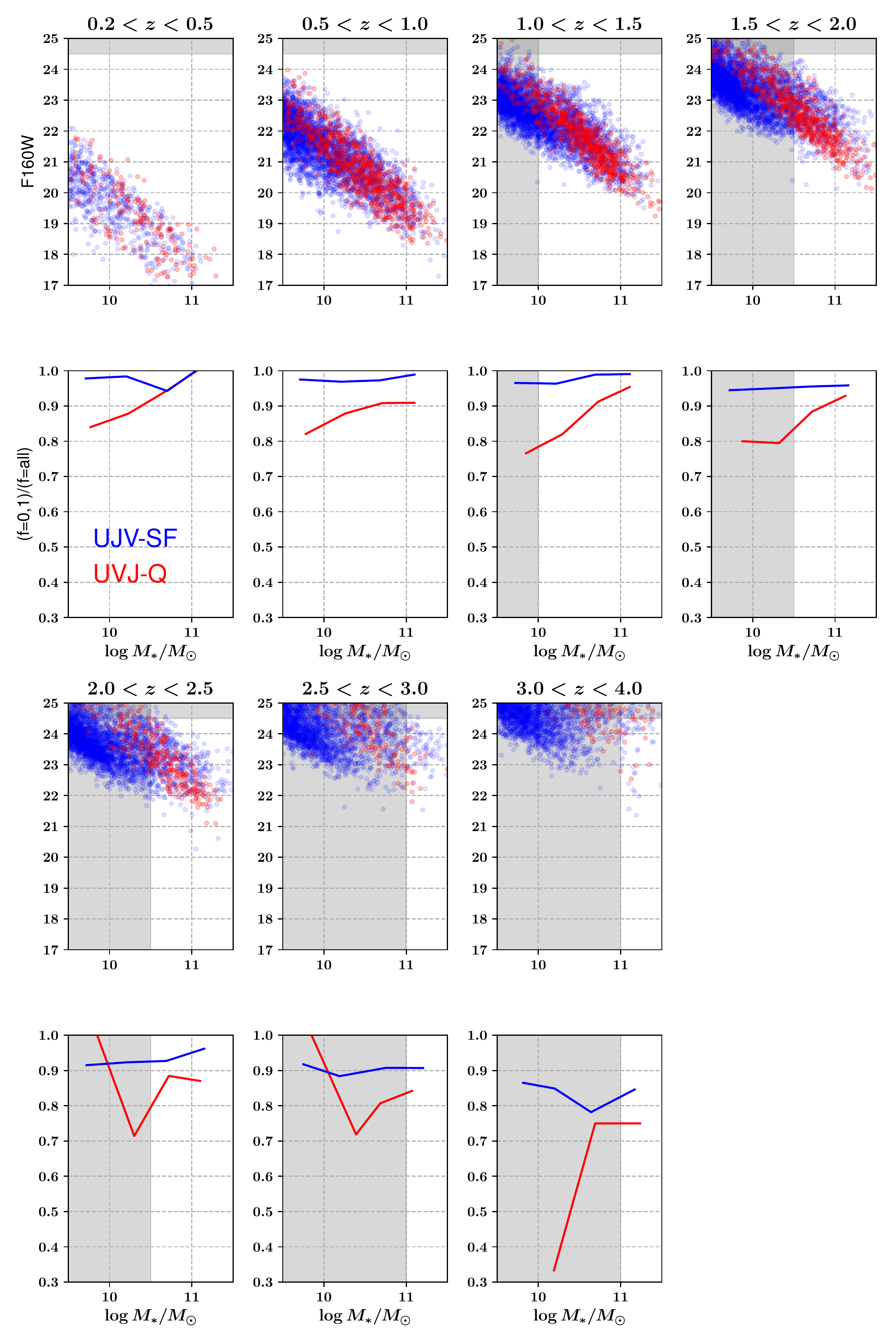}
  \caption{The top panels are the $F160W$ magnitude plotted against mass, with each panel showing a different redshift bin. The bottom panels show the corresponding recovery of `good' fits (i.e. a flag value of 0 or 1 the \citet{vanderwel2012} catalogs) as a function of UVJ star-forming state. The grey-shaded region marks our selected mass and magnitude completeness limits for this study, with the mass limit evolving with increasing $z$.}
  \label{fig:mass_complete}
  \end{center}
\end{figure*}

\section{Sample Selection}

This work makes use of the structural parameter catalogues of
\citet{vanderwel2012} which were generated using \texttt{GALFIT}
\citet{peng2010}. We use the parameters in the observed F160W band,
which corresponds to the H-band. 
{\new These authors constructed PSFs in a hybrid way: the outskirts of the
PSFs are derived from stacked stars in the image; the area within a
radius of 3 pixels is based on theoretical
PSFs constructed by TinyTim \citet{krist1995} and
processed in the same way as the raw science data.
GALFIT is used to fit to each individual galaxy. Neighboring objects
are masked out if they are substantially fainter than the main target,
otherwise they are included in the fit \citep[see][for more details]{vanderwel2012}.
}

We also utilize the most recent
($v4.1.5$) photometric catalogues on which they are based from the
CANDELS/3D-HST survey \citep{brammer2012a, skelton2014,
  momcheva2016}. We use the stellar population parameters, and
rest-frame colours based on the `zbest' catalogues, which will use (if
available) first a spectroscopic redshift, then a (good) grism
redshift and lastly a photometric redshift if a spectroscopic and
grism redshift were not available. 
{\new Stellar masses were estimated from fits of stellar population models
to the full photometric
dataset (ranging from the UV to 4.5 $\mu$m). }
We refer the reader to the
aforementioned papers and their associated documentation for
details. \footnote{https://3dhst.research.yale.edu/Data.php}

We perform a first pass selection using the 3DHST photometric flags ($\mathrm{use\_phot}=1$), as well as an $F160W$ magnitude cut of $m_{AB}=24.5$ to ensure uncertainties in size and shape were within $10\%$ (as described in \citealt{vanderwel2012}). We use objects with a quality flag of $f=0,1$ in \citet{vanderwel2012} which means that \texttt{GALFIT} converged on a solution (without crashing) and that the solution did not require parameters to take on their `constraint' values. 

We also separate our sample into SF and Q galaxies based on their rest-frame $U-V$ and $V-J$ colours, where galaxies display a colour bi-modality and separate based on specific star formation rates \citep{labbe2005, williams2009, williams2010, whitaker2011}. We use the $UVJ$ boundaries defined in \citet{muzzin2013b} to separate the Q and SF sequences. 

In Figure~\ref{fig:mass_complete} we have plotted the $F160W$ AB magnitude, and the fraction of `good' structural fits ($f=0,1$ in \citet{vanderwel2012}) as a function of mass and redshift, as well as SF state to determine our mass completeness as a result of our magnitude limit and the effect of our decision to take only `good' structural parameters. In the top panels we have indicated the mass completeness limit for each redshift (which ranges from $\log{M_{*}/M_{\odot}}=9.5-11.0$), to ensure sufficient signal-to-noise (S/N). In the bottom panels, we see the fraction of `good' structural fits using our mass and magnitude selection is always greater in the SF galaxies, likely because of the difference in their rest-frame optical colours. This is particularly striking for quiescent galaxies at the highest redshift bin ($3.0<z<4.0$) at $\log{M_{*}/M_{\odot}}<10.5$ where we see the recovery of `good' fits is $\sim30\%$.  However, our mass cut from the top panels ensures we have recovered $>80\%$ of the total galaxies in each redshift bin. 

After applying all the aforementioned selection criteria to the complete 3DHST catalogue,  we are left with 9301 galaxies. A census of these galaxies broken down into their respective redshift and UVJ-SF state can be found in Table~\ref{tab:number}. 


\begin{deluxetable}{lcc}
\tabletypesize{\scriptsize}
\tablecaption{Number of galaxies in each redshift range by UVJ SF-state}
\tablewidth{0pt}
\tablehead{
\colhead{$z$-range} & \colhead{$Quiescent$} & \colhead{$Starforming$} 
}
\startdata
$0.2<z<0.5$ & 173 & 589 \\
$0.5<z<1.0$ & 781  & 3426 \\
$1.0<z<1.5$ & 643 & 1904 \\
$1.5<z<2.0$ & 357 & 614 \\
$2.0<z<2.5$ & 187 & 477 \\
$2.5<z<3.0$ & 16 & 78 \\
$3.0<z<4.0$ & 12 & 44
\enddata
\tablecomments{Above are the number of galaxies in each redshift range that are above our mass limits outlined in Fig.~\ref{fig:mass_complete}.}
\label{tab:number}
\end{deluxetable}


\section{Analysis}

\begin{figure*}[ht]
  \begin{center}
  \includegraphics[width=\textwidth]{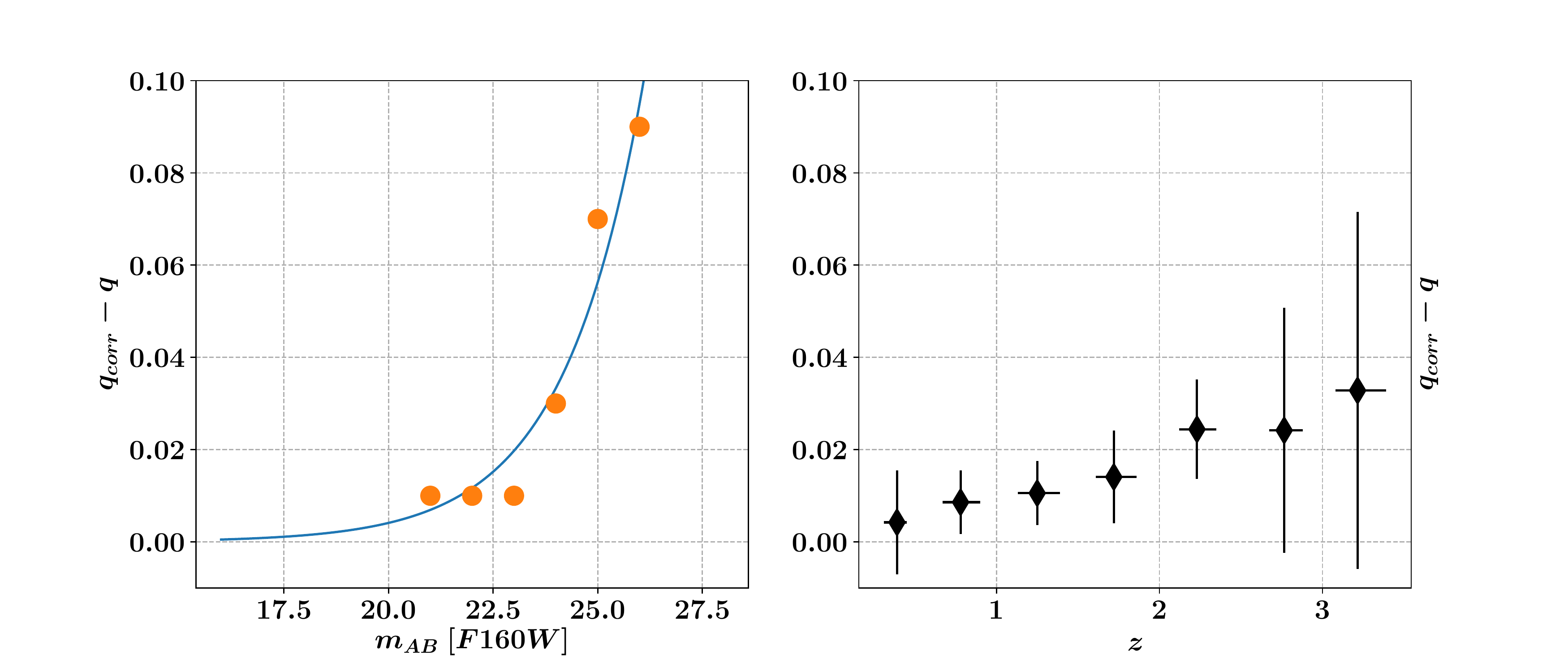}

  \caption{Left: {\new The systematic error in $q$ as measured by
      \citet{vanderwel2012} (orange points; see their Table 3) as a
      function of the $F160W$ magnitude. $q_{corr}$ is the flattening after
      correction for the systematic error. The blue line is an
      exponential fit to the data. Right: The average systematic error in
      $q_{med}$ from the original structural catalog from
      \citet{vanderwel2012} as a function of $z$. The individual
      galaxies have been corrected by using the fit from the left panel.
      The
      error bars show the variance in values. As expected, the total
      effect of the systematics grows bigger with redshift.}}

  \label{fig:systematics}
  \end{center}
\end{figure*}

\subsection{Correcting for Systematics}

Since we are taking a median of $P(q)$, and we have already imposed a fairly conservative S/N cut, our random errors on the median are a fraction of a percent for most data points in this article. However, the systematics in $q$ can be significant at the faintest magnitudes. Since we wish to investigate the trends with flattening out to significant $z$, rather than exclude these galaxies from our sample, we chose to correct for the systematics investigated by \citet{vanderwel2012}. 

In their article, \citet{vanderwel2012} used model light profiles
convolved with the noise and PSF profiles of HST to estimate the
effects of systematics. They repeated their surface brightness profile
fitting on the simulated images and found that near the magnitude
limits of their survey, the measured $q$ in the data were flatter than
the model images. In their Table 3, they tabulated the average
systematic as a function of $F160W$ magnitude, which we have plotted
in Fig.~\ref{fig:systematics}. 
{\new Notice that \citet{vanderwel2012} list ``simulation output - model
  input''
in their Table 3. Hence the correction values shown in Fig. 2 are the
opposite of the listed values, as we show the term that is added to
the observed data.
}
In the left panel of
Fig.~\ref{fig:systematics}, we fit an exponential function to the
data, and made corrections to the values of $q$ in the catalog based
on each object's $F160W$ magnitude. Although we do not know the
magnitude of the systematic for any individual object, our approach of
medians means we can apply these corrections. In the right panel of
Fig.~\ref{fig:systematics} we have shown the median correction as a
function of $z$. As expected, the magnitude of the correction is
larger at higher redshift, where the sample is dominated by objects at
fainter magnitude limit (as seen in the upper panel of
Fig.~\ref{fig:mass_complete}).

Unless otherwise specified, the values of $q$ presented in this paper are corrected for these systematic effects. 

{\new
Another potential systematic can be caused by the shifting intrinsic
bandpass as a function of wavelength.
We tested the effect of bandpass on the axis ratio in two
ways. First, we used the analysis of the GAMA survey \citet{kelvin2012}
.
These authors derived the flattenings in bands ranging from the u band
to the K band. We find that the difference in the median flattening is
very small for this sample. When expressed as a function of
log(wavelength), it is $d \log q_{med}/d \log \lambda$ = 0.00 for quiescent
galaxies and 0.05 for star forming galaxies. This is measured between
the g band and the H band, representative for our sample. The effect
on our results are negligible. In addition, we used the CANDELS
photometry itself to estimate the effect, by comparing the flattening
of the
F125W and F160W bands. We find $d \log q_{med} / d \log \lambda = 0.06 \pm
0.03$ and 0.11 $\pm$ 0.024 for quiescent and star forming galaxies. The 
effect for star
forming galaxies somewhat higher than estimated from GAMA, but
consistent
at the 2.5 $\sigma$ level. It suggests that the dependence of
flattening on passband may depend on redshift. It would still lead to very
small systematics. 
We tested whether this correction would affect our results; and we
only
found a small difference for flattening of the star forming galaxies
as a function of redshift (Fig. 4), where the trend changes by about
0.02 per unit redshift. 
This is a very small trend which will be ignored in the rest of the analysis.
}

\subsection{Trends with star-formation, $M_{*}$, $z$, $r_{e}$ and $n$}

To investigate trends in $q_{med}$ with other properties, we binned our galaxies into 7 different redshift bins (with ranges specified in Table~\ref{tab:number} ), as well as 4 different stellar mass bins ($\log{M_{*}/M_{\odot}}\in$ $[ \,9.5,10.0] \,,$ $ [ \,10.0,10.5] \,,$ $[ \,10.5,11.0] \,, $ $[ \,11.0,12.0] \,$), 3 bins of $r_{e}$ ($r_{e} [kpc] \in$ $[ \,0,3] \,,$ $ [ \,3,6] \,,$ $[ \,6,9] \,, $ $[ \,9,20] \,$), and 3 bins of $n$ ($n\in$$[ \,0,2.0] \,, $ $[ \,2.0,4.0] \,,$ $[ \,4.0,8.0] \,$). We exclude galaxies with $r_{e}<0.1\prime\prime$ from our sample, as this is smaller than the HWHM of the PSF.
{\new The median $q_{med}$ are derived for the various bins, and the errors 
are determined from a bootstrap procedure. Bootstrap
resamples are constructed and the medians are determined. The error
bars shown in the figures are the rms deviations derived from the
distribution of bootstrap medians.
}

\begin{figure*}
  \begin{center}
  \includegraphics[width=\textwidth]{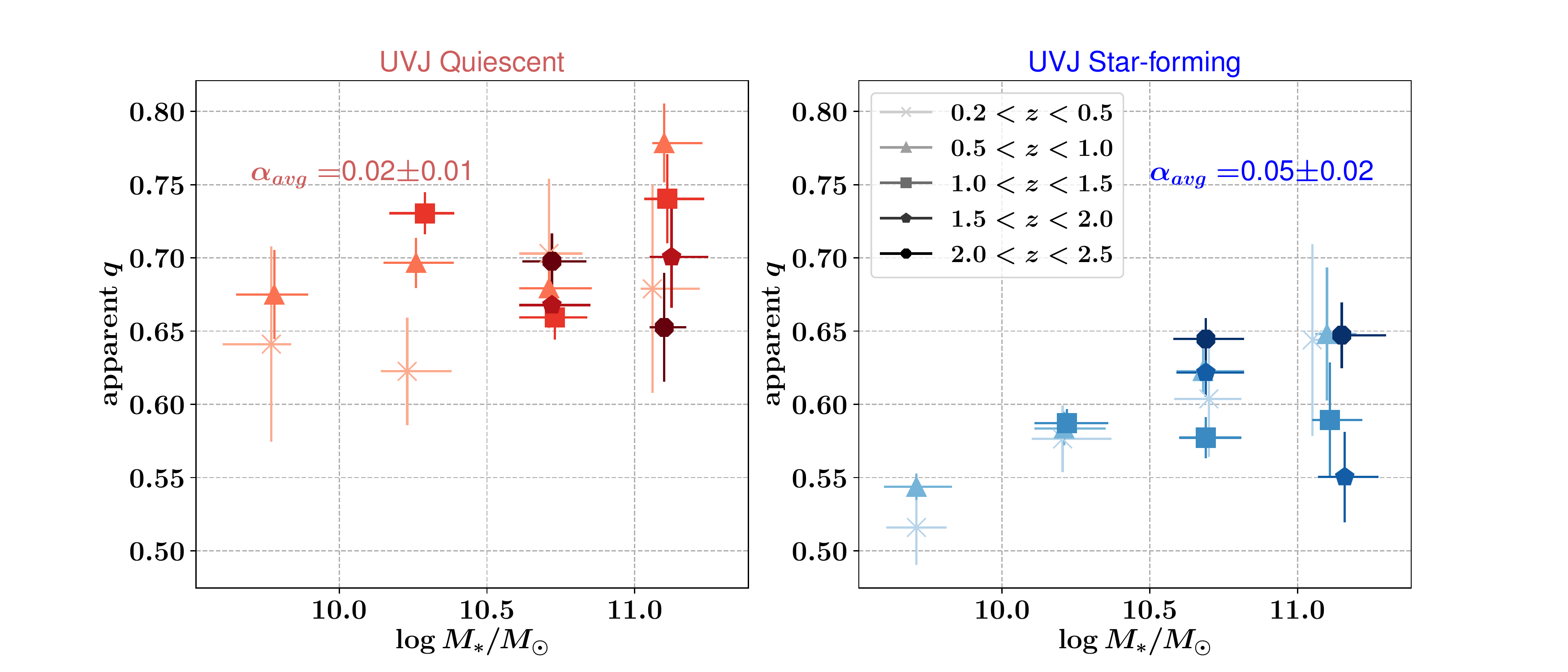}
  \caption{Apparent axis ratio as a function of mass and redshift for
    both UVJ-quiescent (left) and UVJ-SF (right). 
    The quiescent galaxies are rounder than the star forming galaxies;
    and the quiescent galaxies do not show a strong trend with mass.
    On the other hand, the star forming galaxies do show a trend with
    mass: the more massive galaxies are rounder.
    The error bars in
    $\log{M_{*}/M_{\odot}}$ represent the interquartile range, and the
    error bars in $q_{med}$ are the $1\sigma$ range from a
    bootstrapped median, and represents the variance. $\alpha_{avg}$
    is the average best-fit slope for $z<1.5$ (i.e. that is redshift
    ranges that had at least 3 data points).}
  \label{fig:q_by_uvj_mass}
  \end{center}
\end{figure*}

\begin{figure*}
  \begin{center}
  \includegraphics[width=\linewidth]{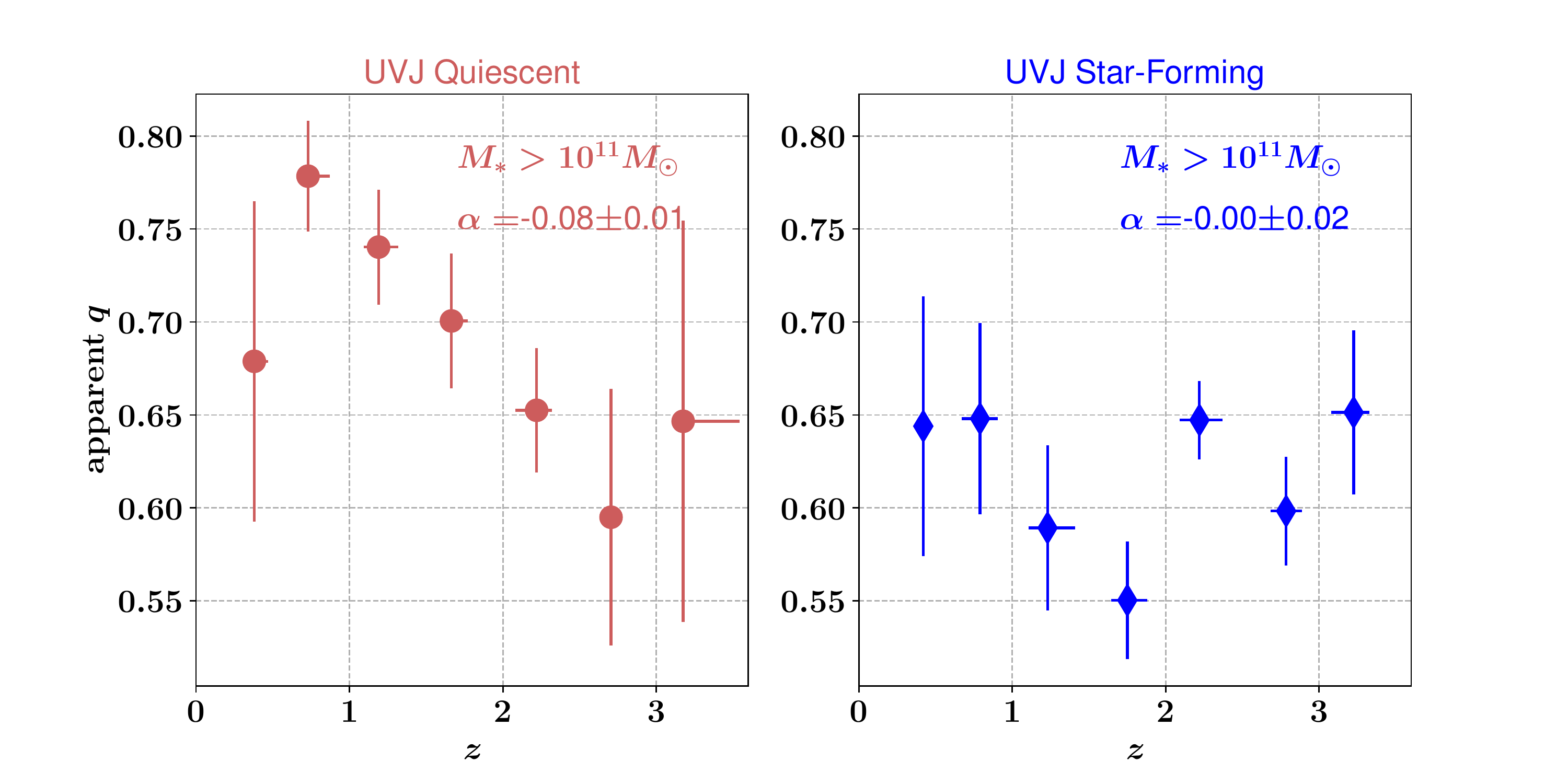}
  \caption{Apparent axis ratio as a function of redshift, and separated into quiescent (left) and star-forming galaxies (right) via a UVJ colour selection for galaxies at $M_{*}>10^{11}M_{\odot}$. The error bars in $q_{med}$ are from the bootstrapped median, and are representative of the scatter, and the error bars in $z$ show the interquartile range. $\alpha$ is the best fit, linear-least squares slope. Here we see the quiescent galaxies are rounder than the star-forming galaxies at $z<2$, but are comparable at $z>2$. We also note that the apparent axis ratio has shown significant evolution in quiescent galaxies, but the trend in $q_{med}$ with $z$ for star forming galaxies is flat.}
  \label{fig:q_z}
  \end{center}
\end{figure*}

In Fig.~\ref{fig:q_by_uvj_mass} we have plotted $q_{med}$ as a function of $\log{M_{*}/M_{\odot}}$ and $z$. In this figure, we only plot our results to $z=2.5$ because we are not complete in mass above this redshift (although we plot our highest mass bin, $M_{*}>10^{11}M_{\odot}$ where we are complete in Fig.~\ref{fig:q_z}). Considering only the quiescent galaxies, we have calculated the average linear least squares slope ($\alpha_{avg}$) for every redshift bin, and find an $\alpha_{avg}=0.01\pm0.01$, which is consistent with $q_{med}$ being independent of $M_{*}$. On the other hand, star-forming galaxies at $z<1$ do display a broad mass dependence, ($\alpha_{avg}=0.05\pm0.02$) with lower mass galaxies appearing flatter than higher mass galaxies. Because we are mass-limited, whether or not this trend continues at $z>1$ is an open question which would require deeper survey depths to answer. 

If we now consider the broad difference between quiescent and star-forming galaxies in Fig.~\ref{fig:q_by_uvj_mass}, we see that the quiescent galaxies are generally rounder than their equivalent mass star-forming counterparts. The exception to this is in our $2.0<z<2.5$ redshift bin, where at $\log{M_{*}/M_{\odot}}>11.0$, the axis ratios are indistinguishable. This could be indicative of similar morphology between the two populations at these redshifts. 

We investigate this similarity to higher redshifts by only considering galaxies in our highest mass bin where we have sufficient redshift coverage given our mass-complete limits. In Fig.~\ref{fig:q_z}, we have plotted the apparent axis ratio of galaxies in our highest mass bin as a function of redshift. We see quiescent galaxies are flatter at higher redshifts of equivalent mass, whereas the star-forming galaxies show little evolution in $q_{med}$ with redshift. As in Fig.~\ref{fig:q_by_uvj_mass}, at $z<2$, the quiescent galaxies are rounder than their star-forming counterparts. At $z>2$, we see that there is no discernible difference in the $q_{med}$ between the star-forming and quiescent populations, suggesting that at this mass (as alluded to in Fig.~\ref{fig:q_by_uvj_mass}) perhaps these galaxies have similar structure. 

\begin{figure*}
  \begin{center}
  \includegraphics[width=\textwidth]{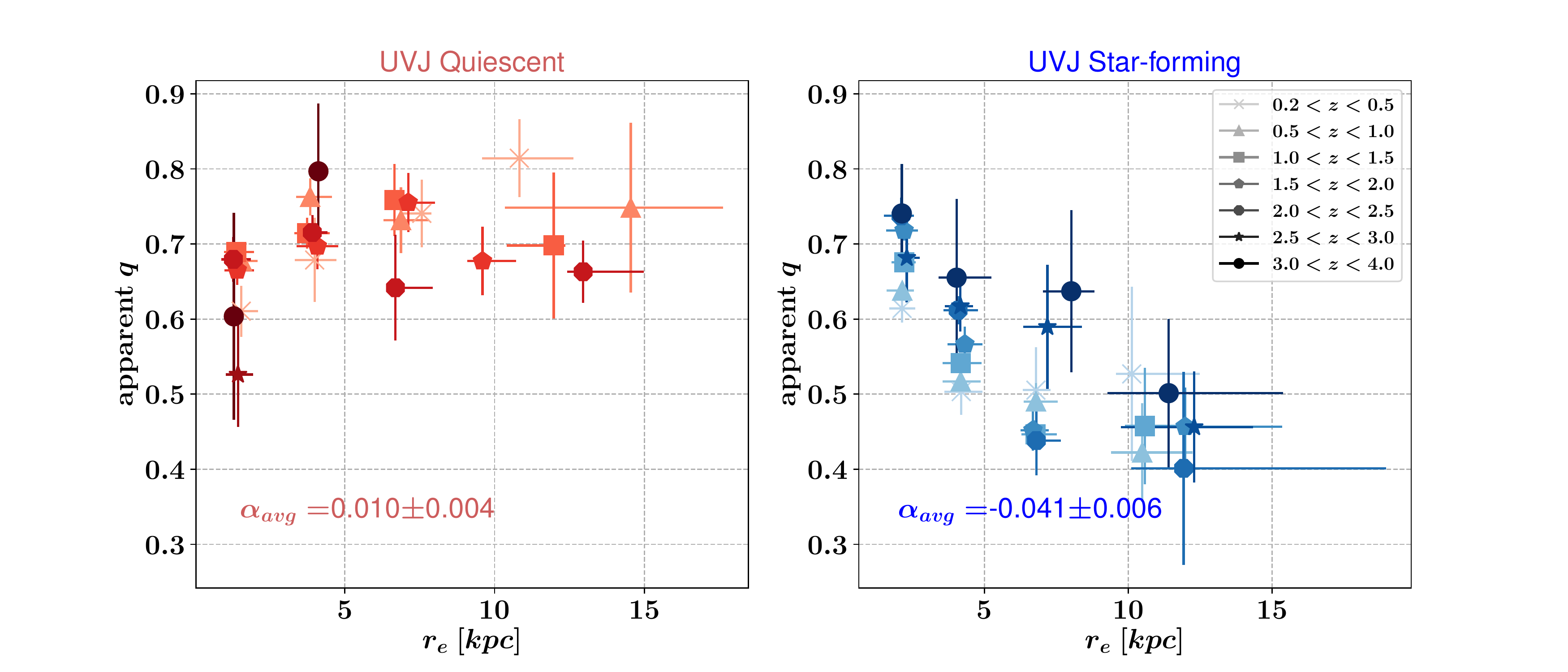}
  \caption{Apparent axis ratio as a function of effective radius for
    the same redshifts bins as Fig. 3. As in previous figures, the
    error bars in $q_{med}$ are the $1\sigma$ from the bootstrapped
    sample, and the errors in $r_{e}$ represent the interquartile
    range. $\alpha_{avg}$ is the average best fit slope for each
    redshift range. As in all other instances in this article, bins
    with 2 or fewer galaxies have been omitted, which is why there are
    missing data points in the left hand panel at $z>2.5$. 
    Quiescent galaxies do not show a strong trend between size and
    flattening; on the other hand, for
    star-forming galaxies, there is a significant anti-correlation
    between $q_{med}$ and $r_{e}$, although no consistent $z$
    evolution. }
  \label{fig:q_by_uvj_re}
  \end{center}
\end{figure*}

\begin{figure*}
  \begin{center}
  \includegraphics[width=\textwidth]{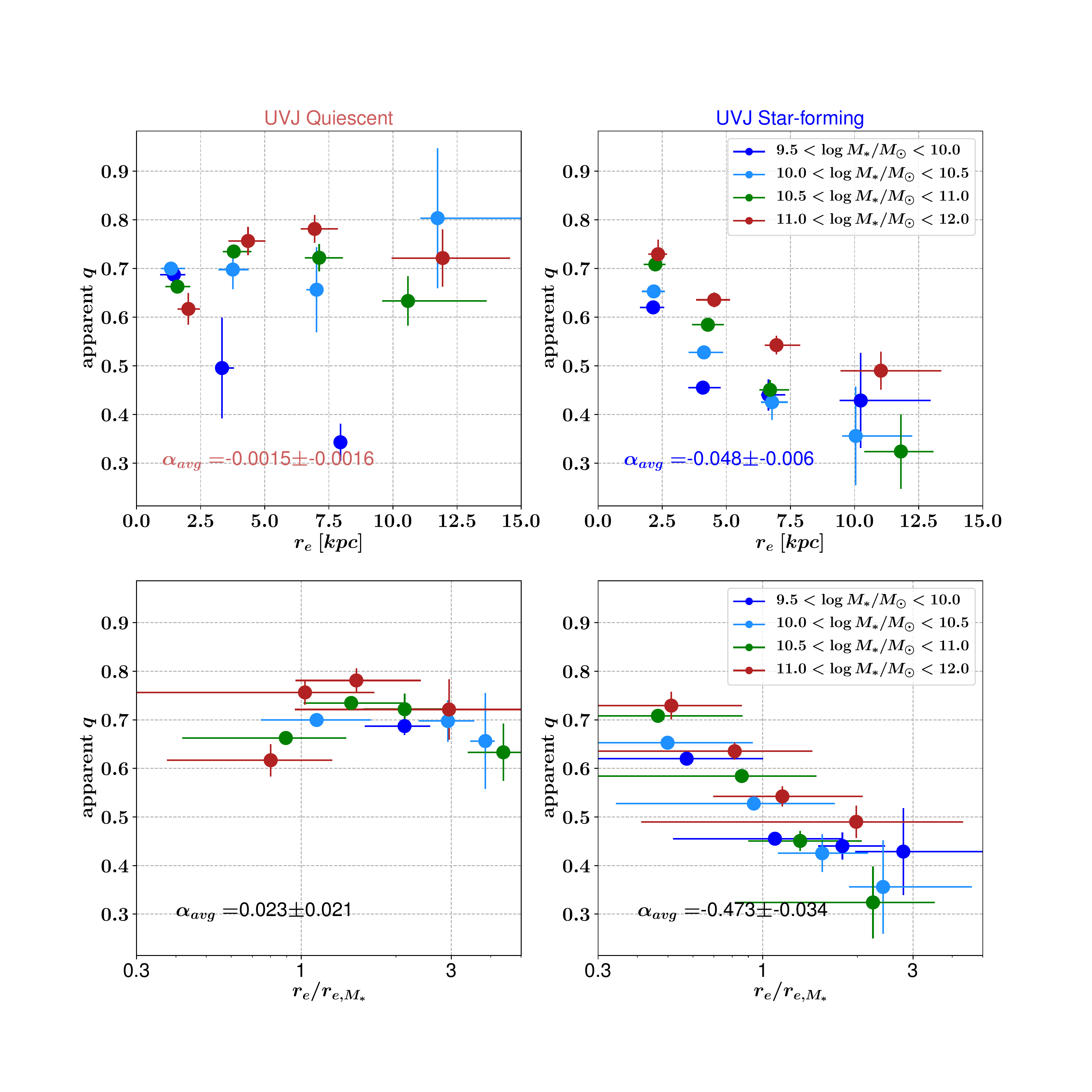}
  \caption{Top: Similar to Fig.~\ref{fig:q_by_uvj_re}, except galaxies
    have been binned according to $M_{*}$ instead of redshift. As in
    other figures, $\alpha_{avg}$ is the average of the best-fit
    linear least squares slopes to each mass bin. Bottom: The same as
    the top row, except instead of plotting the axis ratio against
    $r_{e}$, we have plotted the ratio of $r_{e}$ to the expected size
    based on its mass from the mass-size relation of
    \citet{vanderwel2014a} ($r_{e,M_{*}}$. The differences in mass bin
    seen in the top row disappear when considering the deviation from
    the mass-size relation. We find that the flattening of star
    forming galaxies depends strongly on the size normalized to the
    expected size for the redshift and mass of the galaxies. No strong
  trend is found for quiescent galaxies.}
  \label{fig:q_by_uvj_re_mass}
  \end{center}
\end{figure*}

Given the known association between a galaxy's mass and size
\citep[e.g.,][]{shen2003, vanderwel2014a, lange2015} and that the size
of galaxies at an equivalent mass are observed to be smaller at larger
redshifts \citep[e.g.,][]{daddi2005, trujillo2006, franx2008,
  vandokkum2008, straatman2015}, it is also important to determine
whether the trends observed in Fig.~\ref{fig:q_by_uvj_mass} are driven
by the size evolution. As previously mentioned, we have binned our
data according to $r_{e}$ and have plotted how this evolves with $z$
and $M_{*}$ in Fig.~\ref{fig:q_by_uvj_re} and
Fig.~\ref{fig:q_by_uvj_re_mass}, respectively, but have omitted bins
with fewer than 3 galaxies (as has been done for all medians in this
article).

In Fig.~\ref{fig:q_by_uvj_re}, we see that the $q_{med}$ of star-forming galaxies depends more strongly on $r_{e}$ than their quiescent counterparts (with $\alpha_{avg}=0.01\pm0.004,~-0.039\pm0.007$ for quiescent and star-forming galaxies respectively), with large galaxies being flatter than smaller galaxies. At low-$z$, quiescent galaxies become marginally rounder with increasing size, with this trend disappearing, or even reversing at $z>2$. 

Fig~\ref{fig:q_by_uvj_re_mass} echoes the trends with $r_{e}$ seen in Fig.~\ref{fig:q_by_uvj_re} (with star-forming galaxies showing steeper $\alpha_{avg}$ than quiescent galaxies), however there is a much stronger dependence on $M_{*}$ than with $z$, with massive galaxies always rounder than less massive galaxies at fixed $r_{e}$, with the exception of the smallest quiescent galaxies where the trend reverses. These trends are also what are expected if the B/T ratio increases with increasing $M_{*}$ and decreasing $r_{e}$. In this figure, we also plot $q_{med}$ as a function of $r_{e}$/$r_{e,M_{*}}$, where $r_{e,M_{*}}$ is the expected size given the stellar mass from the mass-size relations of \citet{vanderwel2014a}. This can be thought of as a deviation from the mass-size relation. When plotting this fraction instead of the $r_{e}$, we see the mass dependence largely disappears in both quiescent and star-forming galaxies. In quiescent galaxies we see a relatively flat relationship. For star-forming galaxies, galaxies that lie below the mass-size relation are rounder than those that lie above it. 

\begin{figure*}
  \begin{center}
  \includegraphics[width=\textwidth]{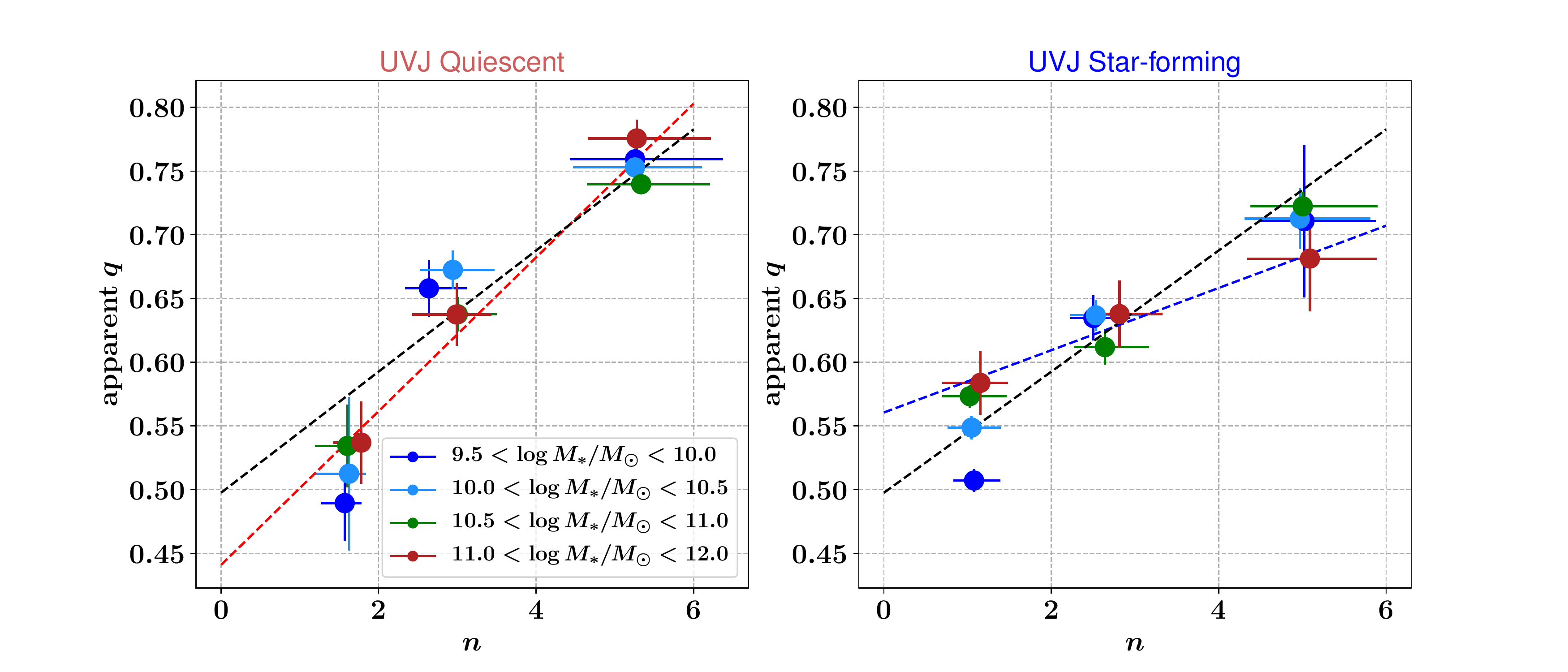}
  \caption{Apparent axis ratio binned by Sersic index for 4 different
    mass bins for UVJ-quiescent (left) and UVJ-SF (right). Black
    dashed linens both panels is the linear least squares fit to the
    combined star-forming and quiescent sample. Red and blue dashed
    lines are the linear fits to the quiescent and star-forming
    galaxies, respectively. In the left panel, we see no apparent mass
    dependence in the quiescent galaxies, but we do see a strong
    correlation of flattening with Sersic index. In the right panel,
    we see a flatter, albeit, still strong relationship between $n$
    and $q_{med}$, with no apparent mass trend, except in the lowest
    mass bin where more massive galaxies are rounder. The slopes are
    $alpha = 0.058, 0.062, 0.035$ for all, quiescent, and star forming
  galaxies respectively.}
  \label{fig:q_n_mass}
  \end{center}
\end{figure*}

In Fig.~\ref{fig:q_n_mass}, we investigate the dependencies of $n$ on $q_{med}$ and $M_{*}$. In this Figure, the galaxies have been binned by $n$. We observe a strong positive correlation between $q_{med}$ and $n$ in both quiescent and star forming galaxies, with no significant $M_{*}$ dependence. Because there is no significant $M_{*}$ dependence, we have plotted trend lines in Fig.~\ref{fig:q_n_mass} based on the median of all galaxies, as well as only the quiescent/star-forming in their respective $n$ bin. These lines show that the $n$ dependence is steeper for quiescent galaxies. This is the most significant trend observed out of the structural parameters investigated. 

\subsection{Is $n$ driving trends with $q_{med}$?}

Because of the tight relationship between $q_{med}$ and $n$, we re-investigate the observed trends with $q_{med}$ to test the extent to which these trends can be explained by trends in $n$. To this end, we re-calculate $q_{med}$ using their measured values of $n$ as well as the relationships for star-forming and quiescent galaxies in Fig.~\ref{fig:q_n_mass}, $q_{n}$. We then take the residual between $q_{med}$ and $q_{n}$ and and plot that against $M_{*}$, $z$ and $r_{e}$.

\begin{figure*}
  \begin{center}
  \includegraphics[width=\textwidth]{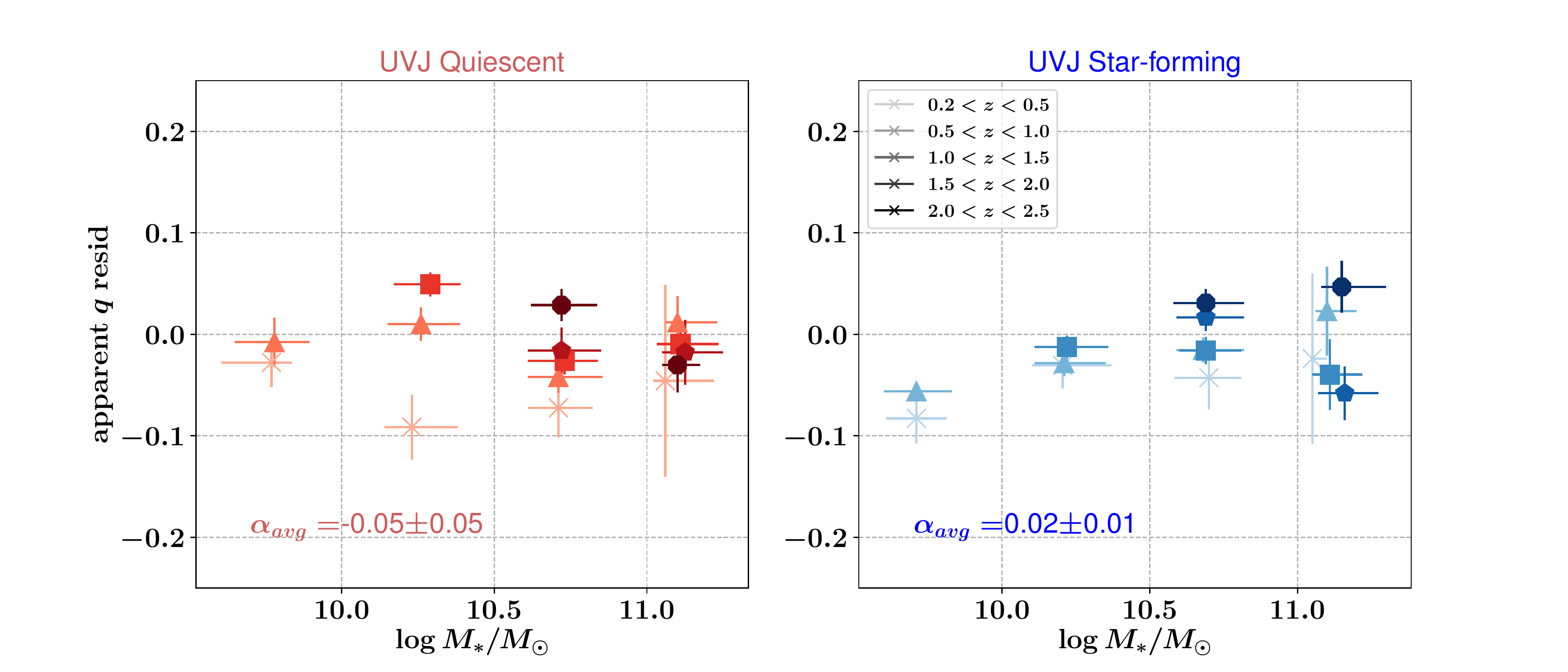}
  \caption{These plots contain the same galaxies and bin as  inFig.~\ref{fig:q_by_uvj_mass}, except the abscissa is now the residual between the actual $q_{med}$, and $q_{n}$ where $q_{n}$ is the $q_{med}$ expected based on the galaxies Sersic index using the quiescent and star-forming relationships from Fig.~\ref{fig:q_n_mass}.  $n$ is able to account for the observed $q_{med}$ to within $\sim10\%$ for most of the mass and redshift bins. Although there is structure in the residuals for the star-forming galaxies, the spread in $q_{med}$ observed in Fig.~\ref{fig:q_by_uvj_mass} disappears, suggesting $n$ is sufficient to explain the trends.}
  \label{fig:q_by_uvj_mass_resid}
  \end{center}
\end{figure*}

Fig.~\ref{fig:q_by_uvj_mass_resid} shows the residuals of the values in Fig.~\ref{fig:q_by_uvj_mass}. In this figure, we see for most data points that the residuals are $\sim10\%$ of the original values, and can account for most of the observed $q_{med}$. For star-forming galaxies, although there is structure in the residuals, $n$ can also account for the trends, especially at the lowest redshifts.

\begin{figure*}
  \begin{center}
  \includegraphics[width=\linewidth]{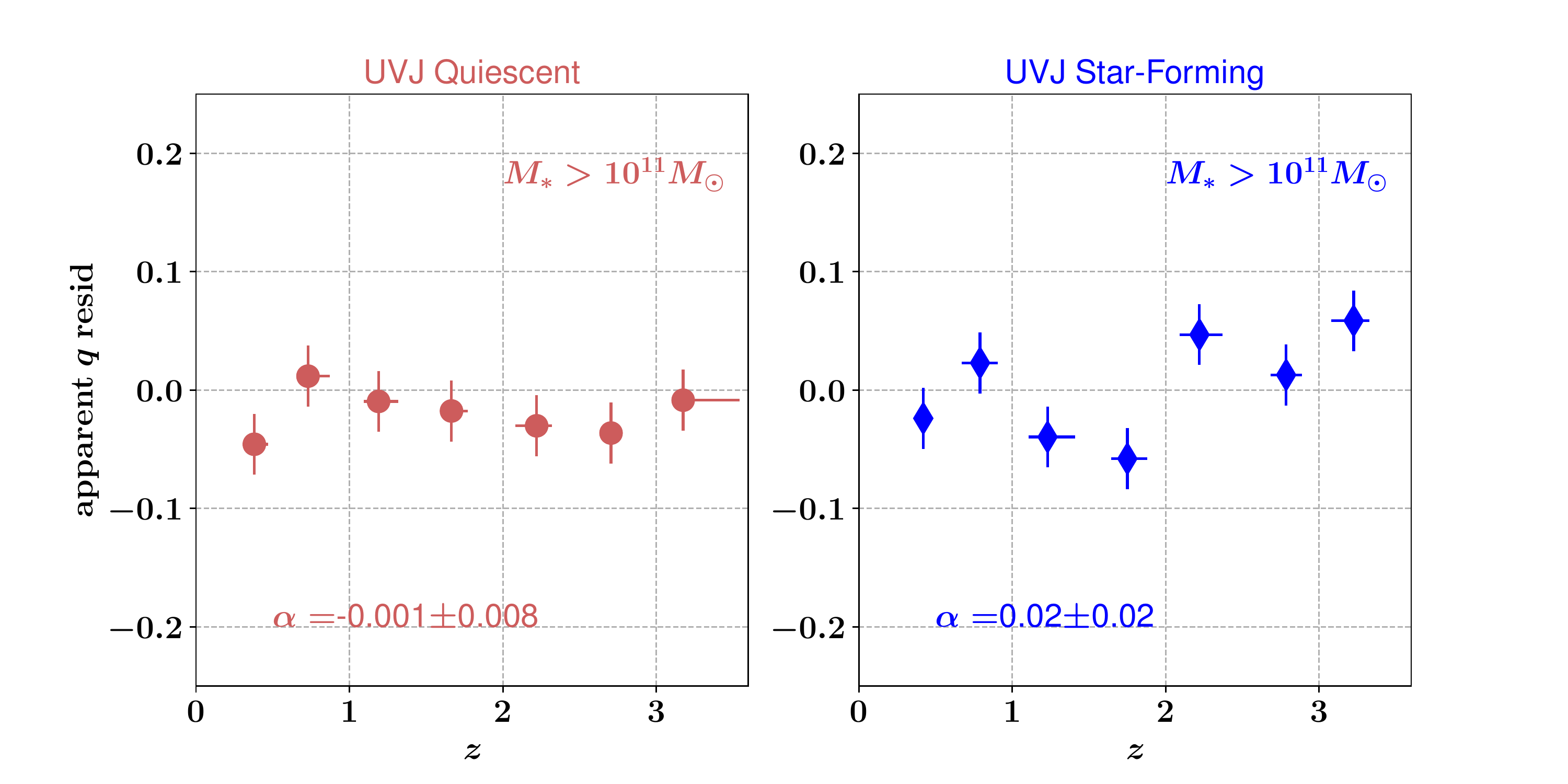}
  \caption{This figure shows the residuals of the relation between
    flattening and redshift for the most massive galaxies
    (Fig.~\ref{fig:q_z}), after subtracting $q_{n}$ (the expected
    $q_{med}$ from a galaxy's $n$ assuming the relationships from
    Fig.~\ref{fig:q_n_mass}) for galaxies at
    $\log{M_{*}/M_{\odot}}>10^{11}$). The strong trend of $q_{med}$
    with $z$ for quiescent galaxies is reduced to zero, showing
    the trend was correlated with a trend with sersic index $n$ and that
    $n$ is able to account for the observed $q_{med}$ for massive
    galaxies. The trend for the star-forming galaxies is still
    consistent with zero.}
  \label{fig:q_z_resid}
  \end{center}
\end{figure*}

In Fig.~\ref{fig:q_z_resid}, we show the residuals of the relationship of our massive galaxy subsample ($M_{*}>10^{11}$) with $z$. In massive galaxies, we see that $q_{med}$ can be fully accounted for by $n$, and the trend of massive galaxies becoming rounder at lower redshift is also gone, with this relationship accounted for by an evolution in the median $n$. We see the flat relationship with star-forming galaxies is also maintained. Therefore, we conclude that the evolution in $n$ can account for any $q_{med}$ evolution in massive galaxies.

\begin{figure*}
  \begin{center}
  \includegraphics[width=\textwidth]{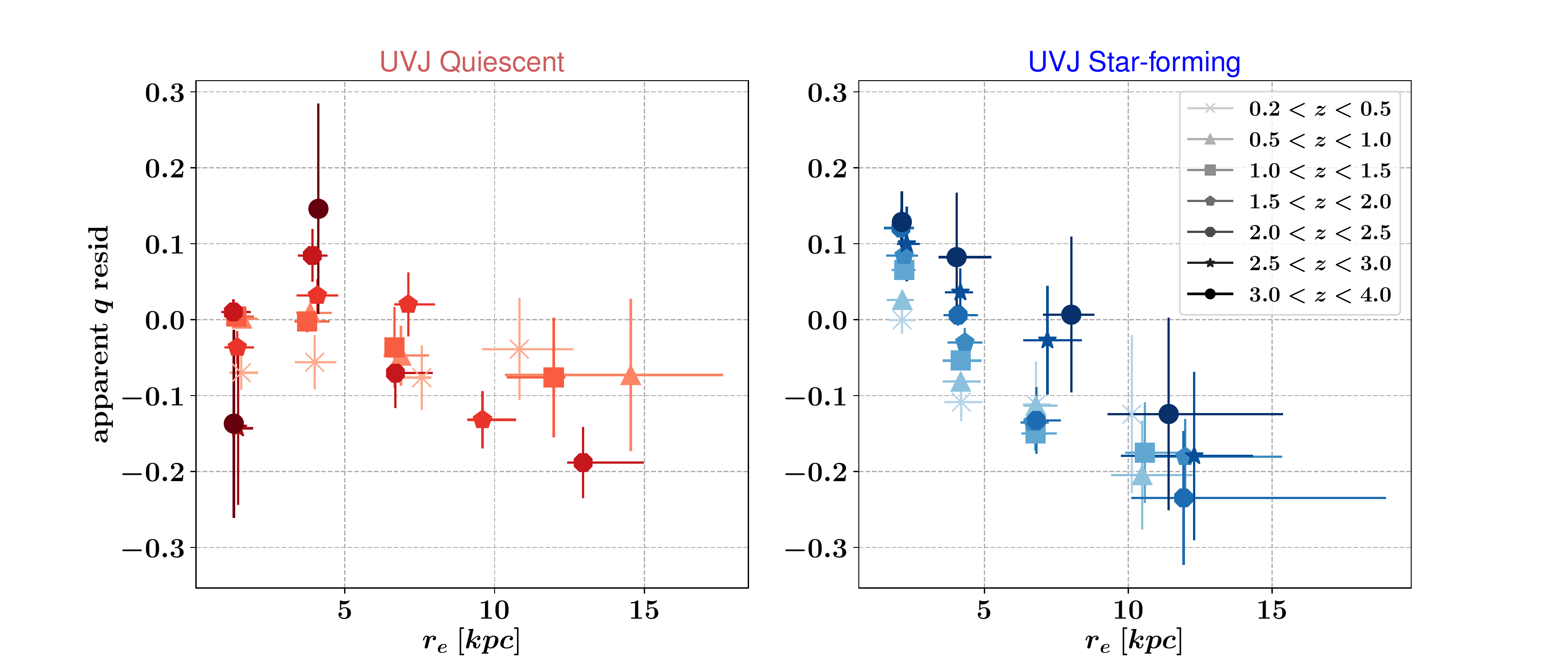}
  \caption{This figure shows the residuals of
    Fig.~\ref{fig:q_by_uvj_re}, after subtracting $q_{n}$ (the
    expected $q_{med}$ from a galaxy's $n$ assuming the relationships
    from Fig.~\ref{fig:q_n_mass}) for galaxies at
    $\log{M_{*}/M_{\odot}}>10^{11}$.  This trend is sufficient to
    explain the observed $q_{med}$ of quiescent galaxies, but does not
    account for the trend of $q_{med}$ with $r_{e}$ in star-forming
    galaxies where the magnitude of the trend persists.}
  \label{fig:q_by_uvj_re_resid}
  \end{center}
\end{figure*}

\begin{figure*}[ht]
  \begin{center}
  \includegraphics[width=\textwidth]{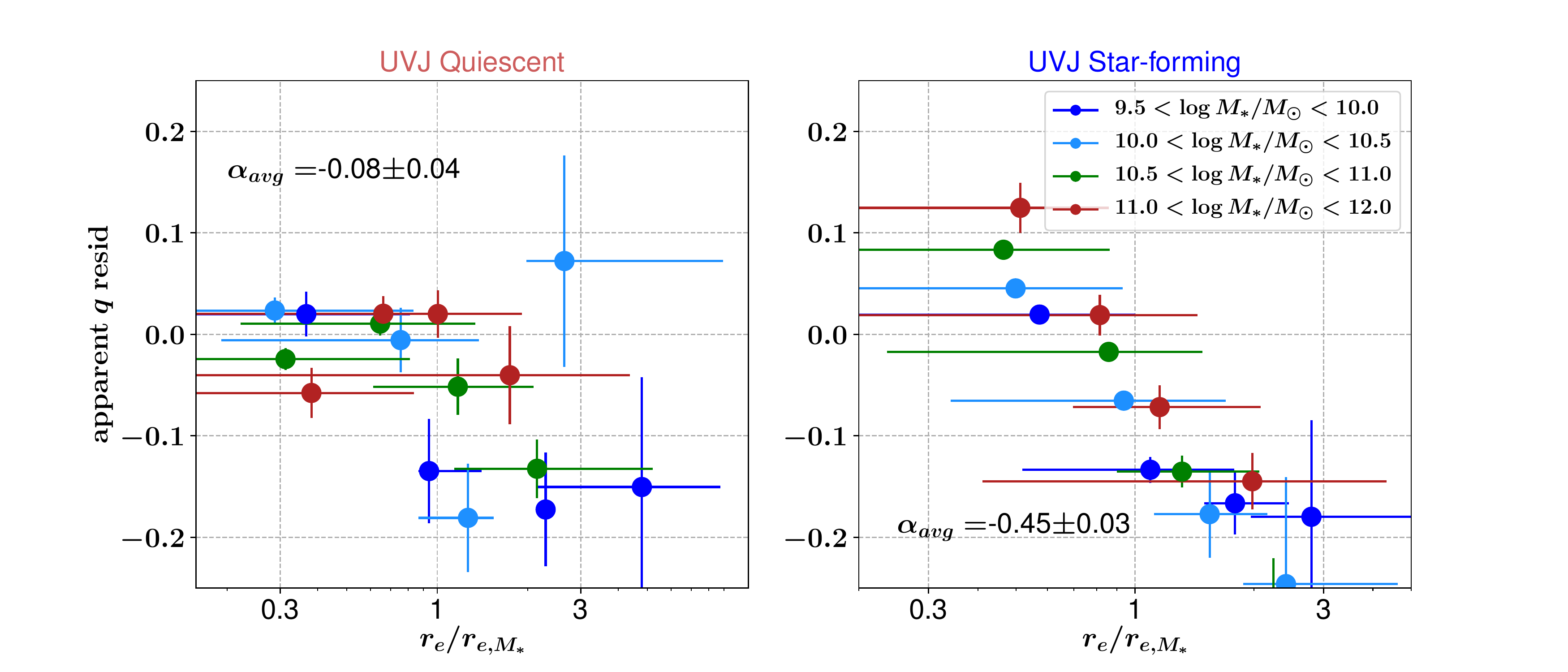}
  \caption{The residual values after subtracting $q_{n}$ (the expected
    $q_{med}$ from a galaxy's $n$ assuming the relationships from
    Fig.~\ref{fig:q_n_mass}) from $q_{med}$ in
    Fig.~\ref{fig:q_by_uvj_re_mass}. As in
    Fig.~\ref{fig:q_by_uvj_re_resid}, $n$ is able to account for
    $\sim80-90\%$ of the observed $q_{med}$ in quiescent galaxies. In
    star-forming galaxies, the observed trend with $r_{e}$ does not
    change after correcting for the correlation with $n$.}
  \label{fig:q_by_uvj_re_mass_resid}
  \end{center}
\end{figure*}

Although $n$ can convincingly account for most of the observed $q_{med}$, as well as trends with $M_{*}$ and $z$, it is insufficient to explain the trends in $r_{e}$ for star-forming galaxies. Fig.~\ref{fig:q_by_uvj_re_resid} and Fig.~\ref{fig:q_by_uvj_re_mass_resid} are the residuals plots of Fig.~\ref{fig:q_by_uvj_re} and Fig.~\ref{fig:q_by_uvj_re_mass}, respectively. For the quiescent galaxies in Fig.~\ref{fig:q_by_uvj_re_mass}, we do see that the previously seen mass dependence of $q_{med}$ at fixed radius is gone (again with the exception of galaxies at the smallest radius). However, the mass dependence for star-forming galaxies persists, as well as the overall trend with $r_{e}$. 

\section{Discussion and Conclusions}

In the previous sections, we investigated the dependence of the
observed $q_{med}$ with various structural parameters. At all masses
below $z<2$, the median quiescent galaxy is rounder than their
star-forming counterpart (Fig~\ref{fig:q_by_uvj_mass}). For quiescent
galaxies, when binned by $M_{*}$ there was no discernible trend with
mass, whereas star-forming galaxies do show a significant mass
dependence at low-redshift ($z<1.0$). {\new At the highest masses
($M_{*}>10^{11}$), quiescent galaxies 
are increasingly flat at
higher $z$, until they match the apparent $q_{med}$ of star-forming
galaxies at $z>2$ }. This
suggests that at the highest redshifts, massive quiescent galaxies are
structurally similar to their star-forming counterparts, and that
high-$z$ quiescent galaxies could be disk-like, a notion that has been
posited previously \citep[e.g.,][]{vanderwel2011, wuyts2011,
  bruce2012, buitrago2013, chang2013, newman2015, hill2017a}

{\new This result is also consistent with studies of nearby relic galaxies,
which are thought to be ``unprocessed'' descendants of high redshift
quiescent galaxies \citep[e.g.][]{vandenbosch2012, trujillo2014, yildirim2017,ferremateu2017}.
}

The observed trend of massive galaxies flattening at higher redshift
(Fig.~\ref{fig:q_z}) can be explained entirely by the dependence of
$n$ on $q_{med}$. This conclusion was drawn through an analysis of the
residuals after subtracting the effect of $n$ from $q_{med}$. To
obtain this correction, we binned our sample according to $n$ and
$M_{*}$ and found $n$ to correlate strongly with $q_{med}$ with no
apparent stellar-mass dependence (Fig.~\ref{fig:q_n_mass}). By using
the linear relationship surmised in Fig.~\ref{fig:q_n_mass}, we
calculated what $q_{med}$ would be given the modelled $n$ from the
catalog of \citet{vanderwel2012}, and plotted the residuals. The
residuals for $q_{med}$ with $z$ in massive galaxies were consistent
with 0 (Fig.~\ref{fig:q_z_resid}), with the conclusion that the
evolution in $n$ drives the evolution in $q_{med}$.

{\new
The $q_{med}$-residuals were also plotted for the other masses, and
the residuals were insignificant for the quiescent galaxies.
These results are consistent with a simple picture in which quiescent
galaxies grow with time due to minor mergers (e.g.,
\citep{vandokkum2010}) which would make them appear rounder and
increase the Sersic index. More detailed comparisons with simulations
are required to test this explanation in detail.

It is remarkable that the star-forming galaxies show different trends
than the quiescent galaxies. This is likely related to the fact that
the star-forming galaxies grow through very different mechanisms
(e.g., growth through the accretion of gas and subsequent star
formation in a disk).

We do not find a strong trend of flattening with redshift (e.g., Fig
4); on the other hand, the flattening correlates significantly with
mass, and very strongly with effective radius; and with Sersic
index. If we ``take out'' the correlation with Sersic index, we still
see a correlation of residual flattening with effective radius, in
contrast to the quiescent galaxies.

The most remarkable of these correlations for star-forming galaxies is
the correlation with $r_e$:} when binning galaxies based on their
$r_{e}$, for star-forming galaxies we observed a negative relationship
between $q_{med}$ and $r_{e}$, with larger galaxies exhibiting
stronger flattening than smaller star forming galaxies, regardless of
$z$ (Fig.~\ref{fig:q_by_uvj_re}). This trend persists when comparing
star-forming galaxies at fixed $r_{e}$ in different mass bins
(Fig.~\ref{fig:q_by_uvj_re_mass}). At fixed $r_{e}$, massive galaxies
are always rounder than lower mass galaxies, regardless of
star-forming state (with the exception of the smallest quiescent
galaxies which requires further investigation). This mass dependence
disappears when considering $q$ as a function from the deviation of
the relevant mass-size relation (Fig.~\ref{fig:q_by_uvj_re_mass}).

{\new
To first order, the results are interpreted by assuming that
$q_{med}$ is
tracing the bulge-to-total galaxy ratio ($B/T$) in star-forming
galaxies. It has been shown previously that $n$ broadly traces $B/T$
in massive galaxies \citep[e.g.,][]{bruce2014b, kennedy2016}; this
combined with the our observation that $q_{med}$ is also correlated
with  $n$ makes a consistent picture.

It is not entirely clear, however, why size plays such a dominant
role: the flattening varies by a factor of about 2 as a function of
size normalized to the mass size relation - stronger than the
variation with Sersic index. In addition, when the dependence on
Sersic index is taken out, there remains a correlation with size.

Possibly, these effects are simply due to the fact that the light
distribution  of star forming galaxies is very sensitive to dust,
orientation, and young, unobscured star formation. Hence simple trends
as for quiescent galaxies become complex - take for example the case
of disk galaxies for which the disks almost ``disappear'' due to dust
when viewed edge-on \citep[e.g.,][]{patel2012}. In short, models are needed to interpret these
results and derive the full interpretation.
}
 
\section{Summary}

We have taken the catalogues of \citet{vanderwel2012} and studied the
evolution of the median apparent axis ratio ($q_{med}$) for over 9000
galaxies out to $z=3$ with $M{*}$, $z$, $n$ and $r_{e}$. We find :

\begin{enumerate}
	\item Quiescent galaxies are rounder than their star-forming
          counterparts at all masses below $z<2$. Above $z>2$, the
          median flattening between massive quiescent and star-forming
          galaxies is identical, suggesting they had very similar
          structure in the early universe {\new (Fig. 4)}. This is an
          extension in redshift of previous work \citep{chang2013} who
          found an increased incidence of disk-like structure in
          massive quiescent galaxies at $z>1$.
	\item The flattening in quiescent galaxies is mass
          independent, whereas in star-forming galaxies, there is a
          steep positive correlation with stellar mass at least until
          $z=1$ {\new (Figs. 3,4)}; due to our mass limits, whether this trend
          continues to higher $z$ is an open question.
	\item In star-forming galaxies, $q_{med}$ correlates
          significantly with $r_{e}$, in contrast to quiescent
          galaxies where there is no discernable trend {\new (Fig. 5)}.
	\item In quiescent galaxies, the strongest common correlation
          was between $q_{med}$ and $n$ {\new (Fig. 7)}. For most relationships, there
          is very little residual correlation between $q_{med}$ and
          $q_{n}$ (the expected $q$ calculated from the s{\'e}rsic
          index), however this was not the case in star-forming
          galaxies {\new (Fig. 8)}.
	\item We suspect that $q_{med}$ is likely tracing the $B/T$
          ratio which would explain why smaller/more massive
          star-forming galaxies are rounder than their extended/less
          massive counterparts, as well as why we do not observe
          strong $M_{*}$ and $r_{e}$ dependencies in quiescent
          galaxies, as the majority of the quiescent galaxies are not
          expected to have prominent disks. We caveat that we are also
          only tracing the light, which would weight blue disks with
          lower mass-to-light ratios heavily in the observables, and
          that the mass distribution could be quite different.
\end{enumerate}

\section{Acknowledgements} 

{\new We thank the referee for the constructive comments which helped to
improve the paper.}
This research has made use of NASA's Astrophysics Data System. This work is based on observations taken by the 3D-HST Treasury Program (GO 12177 and 12328) with the NASA/ESA HST, which is operated by the Association of Universities for Research in Astronomy, Inc., under NASA contract NAS5-26555.
{\new We used the public available programming
language PYTHON, including the NUMPY, MATPLOTLIB packages. }

\bibliographystyle{apj}
\bibliography{lit}

\begin{thebibliography}{44}
\expandafter\ifx\csname natexlab\endcsname\relax\def\natexlab#1{#1}\fi

\bibitem[{{Blanton} {et~al.}(2003){Blanton}, {Hogg}, {Bahcall}, {Baldry},
  {Brinkmann}, {Csabai}, {Eisenstein}, {Fukugita}, {Gunn}, {Ivezi{\'c}},
  {Lamb}, {Lupton}, {Loveday}, {Munn}, {Nichol}, {Okamura}, {Schlegel},
  {Shimasaku}, {Strauss}, {Vogeley}, \& {Weinberg}}]{blanton2003}
{Blanton}, M.~R., {Hogg}, D.~W., {Bahcall}, N.~A., {et~al.} 2003, \apj, 594,
  186

\bibitem[{{Brammer} {et~al.}(2012){Brammer}, {van Dokkum}, {Franx},
  {Fumagalli}, {Patel}, {Rix}, {Skelton}, {Kriek}, {Nelson}, {Schmidt},
  {Bezanson}, {da Cunha}, {Erb}, {Fan}, {F{\"o}rster Schreiber}, {Illingworth},
  {Labb{\'e}}, {Leja}, {Lundgren}, {Magee}, {Marchesini}, {McCarthy},
  {Momcheva}, {Muzzin}, {Quadri}, {Steidel}, {Tal}, {Wake}, {Whitaker}, \&
  {Williams}}]{brammer2012a}
{Brammer}, G.~B., {van Dokkum}, P.~G., {Franx}, M., {et~al.} 2012, \apjs, 200,
  13

\bibitem[{{Bruce} {et~al.}(2012){Bruce}, {Dunlop}, {Cirasuolo}, {McLure},
  {Targett}, {Bell}, {Croton}, {Dekel}, {Faber}, {Ferguson}, {Grogin},
  {Kocevski}, {Koekemoer}, {Koo}, {Lai}, {Lotz}, {McGrath}, {Newman}, \& {van
  der Wel}}]{bruce2012}
{Bruce}, V.~A., {Dunlop}, J.~S., {Cirasuolo}, M., {et~al.} 2012, \mnras, 427,
  1666

\bibitem[{{Bruce} {et~al.}(2014{\natexlab{a}}){Bruce}, {Dunlop}, {McLure},
  {Cirasuolo}, {Buitrago}, {Bowler}, {Targett}, {Bell}, {McIntosh}, {Dekel},
  {Faber}, {Ferguson}, {Grogin}, {Hartley}, {Kocevski}, {Koekemoer}, {Koo}, \&
  {McGrath}}]{bruce2014a}
{Bruce}, V.~A., {Dunlop}, J.~S., {McLure}, R.~J., {et~al.} 2014{\natexlab{a}},
  \mnras, 444, 1001

\bibitem[{{Bruce} {et~al.}(2014{\natexlab{b}}){Bruce}, {Dunlop}, {McLure},
  {Cirasuolo}, {Buitrago}, {Bowler}, {Targett}, {Bell}, {McIntosh}, {Dekel},
  {Faber}, {Ferguson}, {Grogin}, {Hartley}, {Kocevski}, {Koekemoer}, {Koo}, \&
  {McGrath}}]{bruce2014b}
---. 2014{\natexlab{b}}, \mnras, 444, 1660

\bibitem[{{Buitrago} {et~al.}(2013){Buitrago}, {Trujillo}, {Conselice}, \&
  {Haussler}}]{buitrago2013}
{Buitrago}, F., {Trujillo}, I., {Conselice}, C.~J., \& {Haussler}, B. 2013,
  \mnras, 428, 1460

\bibitem[{{Chang} {et~al.}(2013){Chang}, {van der Wel}, {Rix}, {Holden},
  {Bell}, {McGrath}, {Wuyts}, {H{\"a}ussler}, {Barden}, {Faber}, {Mozena},
  {Ferguson}, {Guo}, {Galametz}, {Grogin}, {Kocevski}, {Koekemoer}, {Dekel},
  {Huang}, {Hathi}, \& {Donley}}]{chang2013}
{Chang}, Y.-Y., {van der Wel}, A., {Rix}, H.-W., {et~al.} 2013, \apj, 773, 149

\bibitem[{{Daddi} {et~al.}(2005){Daddi}, {Renzini}, {Pirzkal}, {Cimatti},
  {Malhotra}, {Stiavelli}, {Xu}, {Pasquali}, {Rhoads}, {Brusa}, {di Serego
  Alighieri}, {Ferguson}, {Koekemoer}, {Moustakas}, {Panagia}, \&
  {Windhorst}}]{daddi2005}
{Daddi}, E., {Renzini}, A., {Pirzkal}, N., {et~al.} 2005, \apj, 626, 680

\bibitem[{{Ferr{\'e}-Mateu} {et~al.}(2017){Ferr{\'e}-Mateu}, {Trujillo},
  {Mart{\'{\i}}n-Navarro}, {Vazdekis}, {Mezcua}, {Balcells}, \&
  {Dom{\'{\i}}nguez}}]{ferremateu2017}
{Ferr{\'e}-Mateu}, A., {Trujillo}, I., {Mart{\'{\i}}n-Navarro}, I., {et~al.}
  2017, \mnras, 467, 1929

\bibitem[{{Franx} {et~al.}(1991){Franx}, {Illingworth}, \& {de
  Zeeuw}}]{franx1991}
{Franx}, M., {Illingworth}, G., \& {de Zeeuw}, T. 1991, \apj, 383, 112

\bibitem[{{Franx} {et~al.}(2008){Franx}, {van Dokkum}, {F{\"o}rster Schreiber},
  {Wuyts}, {Labb{\'e}}, \& {Toft}}]{franx2008}
{Franx}, M., {van Dokkum}, P.~G., {F{\"o}rster Schreiber}, N.~M., {et~al.}
  2008, \apj, 688, 770

\bibitem[{{Hill} {et~al.}(2017){Hill}, {Muzzin}, {Franx}, {Clauwens},
  {Schreiber}, {Marchesini}, {Stefanon}, {Labbe}, {Brammer}, {Caputi}, {Fynbo},
  {Milvang-Jensen}, {Skelton}, {van Dokkum}, \& {Whitaker}}]{hill2017a}
{Hill}, A.~R., {Muzzin}, A., {Franx}, M., {et~al.} 2017, \apj, 837, 147

\bibitem[{{Kelvin} {et~al.}(2012){Kelvin}, {Driver}, {Robotham}, {Hill},
  {Alpaslan}, {Baldry}, {Bamford}, {Bland-Hawthorn}, {Brough}, {Graham},
  {H{\"a}ussler}, {Hopkins}, {Liske}, {Loveday}, {Norberg}, {Phillipps},
  {Popescu}, {Prescott}, {Taylor}, \& {Tuffs}}]{kelvin2012}
{Kelvin}, L.~S., {Driver}, S.~P., {Robotham}, A.~S.~G., {et~al.} 2012, \mnras,
  421, 1007

\bibitem[{{Kennedy} {et~al.}(2016){Kennedy}, {Bamford}, {H{\"a}u{\ss}ler},
  {Baldry}, {Bremer}, {Brough}, {Brown}, {Driver}, {Duncan}, {Graham},
  {Holwerda}, {Hopkins}, {Kelvin}, {Lange}, {Phillipps}, {Vika}, \&
  {Vulcani}}]{kennedy2016}
{Kennedy}, R., {Bamford}, S.~P., {H{\"a}u{\ss}ler}, B., {et~al.} 2016, \mnras,
  460, 3458

\bibitem[{{Krist}(1995)}]{krist1995}
{Krist}, J. 1995, in Astronomical Society of the Pacific Conference Series,
  Vol.~77, Astronomical Data Analysis Software and Systems IV, ed. R.~A.
  {Shaw}, H.~E. {Payne}, \& J.~J.~E. {Hayes}, 349

\bibitem[{{Labb{\'e}} {et~al.}(2005){Labb{\'e}}, {Huang}, {Franx}, {Rudnick},
  {Barmby}, {Daddi}, {van Dokkum}, {Fazio}, {Schreiber}, {Moorwood}, {Rix},
  {R{\"o}ttgering}, {Trujillo}, \& {van der Werf}}]{labbe2005}
{Labb{\'e}}, I., {Huang}, J., {Franx}, M., {et~al.} 2005, \apjl, 624, L81

\bibitem[{{Lambas} {et~al.}(1992){Lambas}, {Maddox}, \& {Loveday}}]{lambas1992}
{Lambas}, D.~G., {Maddox}, S.~J., \& {Loveday}, J. 1992, \mnras, 258, 404

\bibitem[{{Lange} {et~al.}(2015){Lange}, {Driver}, {Robotham}, {Kelvin},
  {Graham}, {Alpaslan}, {Andrews}, {Baldry}, {Bamford}, {Bland-Hawthorn},
  {Brough}, {Cluver}, {Conselice}, {Davies}, {Haeussler}, {Konstantopoulos},
  {Loveday}, {Moffett}, {Norberg}, {Phillipps}, {Taylor},
  {L{\'o}pez-S{\'a}nchez}, \& {Wilkins}}]{lange2015}
{Lange}, R., {Driver}, S.~P., {Robotham}, A.~S.~G., {et~al.} 2015, \mnras, 447,
  2603

\bibitem[{{Law} {et~al.}(2012){Law}, {Steidel}, {Shapley}, {Nagy}, {Reddy}, \&
  {Erb}}]{law2012}
{Law}, D.~R., {Steidel}, C.~C., {Shapley}, A.~E., {et~al.} 2012, \apj, 745, 85

\bibitem[{{Momcheva} {et~al.}(2016){Momcheva}, {Brammer}, {van Dokkum},
  {Skelton}, {Whitaker}, {Nelson}, {Fumagalli}, {Maseda}, {Leja}, {Franx},
  {Rix}, {Bezanson}, {Da Cunha}, {Dickey}, {F{\"o}rster Schreiber},
  {Illingworth}, {Kriek}, {Labb{\'e}}, {Ulf Lange}, {Lundgren}, {Magee},
  {Marchesini}, {Oesch}, {Pacifici}, {Patel}, {Price}, {Tal}, {Wake}, {van der
  Wel}, \& {Wuyts}}]{momcheva2016}
{Momcheva}, I.~G., {Brammer}, G.~B., {van Dokkum}, P.~G., {et~al.} 2016, \apjs,
  225, 27

\bibitem[{{Muzzin} {et~al.}(2013){Muzzin}, {Marchesini}, {Stefanon}, {Franx},
  {McCracken}, {Milvang-Jensen}, {Dunlop}, {Fynbo}, {Brammer}, {Labb{\'e}}, \&
  {van Dokkum}}]{muzzin2013b}
{Muzzin}, A., {Marchesini}, D., {Stefanon}, M., {et~al.} 2013, \apj, 777, 18

\bibitem[{{Newman} {et~al.}(2015){Newman}, {Belli}, \& {Ellis}}]{newman2015}
{Newman}, A.~B., {Belli}, S., \& {Ellis}, R.~S. 2015, \apjl, 813, L7

\bibitem[{{Patel} {et~al.}(2012){Patel}, {Holden}, {Kelson}, {Franx}, {van der
  Wel}, \& {Illingworth}}]{patel2012}
{Patel}, S.~G., {Holden}, B.~P., {Kelson}, D.~D., {et~al.} 2012, \apjl, 748,
  L27

\bibitem[{{Peng} {et~al.}(2010){Peng}, {Ho}, {Impey}, \& {Rix}}]{peng2010}
{Peng}, C.~Y., {Ho}, L.~C., {Impey}, C.~D., \& {Rix}, H.-W. 2010, \aj, 139,
  2097

\bibitem[{{Roberts} \& {Haynes}(1994)}]{roberts1994}
{Roberts}, M.~S., \& {Haynes}, M.~P. 1994, \araa, 32, 115

\bibitem[{{Sandage} {et~al.}(1970){Sandage}, {Freeman}, \&
  {Stokes}}]{sandage1970}
{Sandage}, A., {Freeman}, K.~C., \& {Stokes}, N.~R. 1970, \apj, 160, 831

\bibitem[{{Shen} {et~al.}(2003){Shen}, {Mo}, {White}, {Blanton}, {Kauffmann},
  {Voges}, {Brinkmann}, \& {Csabai}}]{shen2003}
{Shen}, S., {Mo}, H.~J., {White}, S.~D.~M., {et~al.} 2003, \mnras, 343, 978

\bibitem[{{Simons} {et~al.}(2017){Simons}, {Kassin}, {Weiner}, {Faber},
  {Trump}, {Heckman}, {Koo}, {Pacifici}, {Primack}, {Snyder}, \& {de la
  Vega}}]{simons2017a}
{Simons}, R.~C., {Kassin}, S.~A., {Weiner}, B.~J., {et~al.} 2017, \apj, 843, 46

\bibitem[{{Skelton} {et~al.}(2014){Skelton}, {Whitaker}, {Momcheva}, {Brammer},
  {van Dokkum}, {Labb{\'e}}, {Franx}, {van der Wel}, {Bezanson}, {Da Cunha},
  {Fumagalli}, {F{\"o}rster Schreiber}, {Kriek}, {Leja}, {Lundgren}, {Magee},
  {Marchesini}, {Maseda}, {Nelson}, {Oesch}, {Pacifici}, {Patel}, {Price},
  {Rix}, {Tal}, {Wake}, \& {Wuyts}}]{skelton2014}
{Skelton}, R.~E., {Whitaker}, K.~E., {Momcheva}, I.~G., {et~al.} 2014, \apjs,
  214, 24

\bibitem[{{Straatman} {et~al.}(2015){Straatman}, {Labb{\'e}}, {Spitler},
  {Glazebrook}, {Tomczak}, {Allen}, {Brammer}, {Cowley}, {van Dokkum},
  {Kacprzak}, {Kawinwanichakij}, {Mehrtens}, {Nanayakkara}, {Papovich},
  {Persson}, {Quadri}, {Rees}, {Tilvi}, {Tran}, \& {Whitaker}}]{straatman2015}
{Straatman}, C.~M.~S., {Labb{\'e}}, I., {Spitler}, L.~R., {et~al.} 2015, \apjl,
  808, L29

\bibitem[{{Trujillo} {et~al.}(2014){Trujillo}, {Ferr{\'e}-Mateu}, {Balcells},
  {Vazdekis}, \& {S{\'a}nchez-Bl{\'a}zquez}}]{trujillo2014}
{Trujillo}, I., {Ferr{\'e}-Mateu}, A., {Balcells}, M., {Vazdekis}, A., \&
  {S{\'a}nchez-Bl{\'a}zquez}, P. 2014, \apjl, 780, L20

\bibitem[{{Trujillo} {et~al.}(2006){Trujillo}, {F{\"o}rster Schreiber},
  {Rudnick}, {Barden}, {Franx}, {Rix}, {Caldwell}, {McIntosh}, {Toft},
  {H{\"a}ussler}, {Zirm}, {van Dokkum}, {Labb{\'e}}, {Moorwood},
  {R{\"o}ttgering}, {van der Wel}, {van der Werf}, \& {van
  Starkenburg}}]{trujillo2006}
{Trujillo}, I., {F{\"o}rster Schreiber}, N.~M., {Rudnick}, G., {et~al.} 2006,
  \apj, 650, 18

\bibitem[{{van den Bosch} {et~al.}(2012){van den Bosch}, {Gebhardt},
  {G{\"u}ltekin}, {van de Ven}, {van der Wel}, \& {Walsh}}]{vandenbosch2012}
{van den Bosch}, R. C.~E., {Gebhardt}, K., {G{\"u}ltekin}, K., {et~al.} 2012,
  \nat, 491, 729

\bibitem[{{van der Wel} {et~al.}(2011){van der Wel}, {Rix}, {Wuyts}, {McGrath},
  {Koekemoer}, {Bell}, {Holden}, {Robaina}, \& {McIntosh}}]{vanderwel2011}
{van der Wel}, A., {Rix}, H.-W., {Wuyts}, S., {et~al.} 2011, \apj, 730, 38

\bibitem[{{van der Wel} {et~al.}(2012){van der Wel}, {Bell}, {H{\"a}ussler},
  {McGrath}, {Chang}, {Guo}, {McIntosh}, {Rix}, {Barden}, {Cheung}, {Faber},
  {Ferguson}, {Galametz}, {Grogin}, {Hartley}, {Kartaltepe}, {Kocevski},
  {Koekemoer}, {Lotz}, {Mozena}, {Peth}, \& {Peng}}]{vanderwel2012}
{van der Wel}, A., {Bell}, E.~F., {H{\"a}ussler}, B., {et~al.} 2012, \apjs,
  203, 24

\bibitem[{{van der Wel} {et~al.}(2014{\natexlab{a}}){van der Wel}, {Franx},
  {van Dokkum}, {Skelton}, {Momcheva}, {Whitaker}, {Brammer}, {Bell}, {Rix},
  {Wuyts}, {Ferguson}, {Holden}, {Barro}, {Koekemoer}, {Chang}, {McGrath},
  {H{\"a}ussler}, {Dekel}, {Behroozi}, {Fumagalli}, {Leja}, {Lundgren},
  {Maseda}, {Nelson}, {Wake}, {Patel}, {Labb{\'e}}, {Faber}, {Grogin}, \&
  {Kocevski}}]{vanderwel2014a}
{van der Wel}, A., {Franx}, M., {van Dokkum}, P.~G., {et~al.}
  2014{\natexlab{a}}, \apj, 788, 28

\bibitem[{{van der Wel} {et~al.}(2014{\natexlab{b}}){van der Wel}, {Chang},
  {Bell}, {Holden}, {Ferguson}, {Giavalisco}, {Rix}, {Skelton}, {Whitaker},
  {Momcheva}, {Brammer}, {Kassin}, {Martig}, {Dekel}, {Ceverino}, {Koo},
  {Mozena}, {van Dokkum}, {Franx}, {Faber}, \& {Primack}}]{vanderwel2014b}
{van der Wel}, A., {Chang}, Y.-Y., {Bell}, E.~F., {et~al.} 2014{\natexlab{b}},
  \apjl, 792, L6

\bibitem[{{van Dokkum} {et~al.}(2008){van Dokkum}, {Franx}, {Kriek}, {Holden},
  {Illingworth}, {Magee}, {Bouwens}, {Marchesini}, {Quadri}, {Rudnick},
  {Taylor}, \& {Toft}}]{vandokkum2008}
{van Dokkum}, P.~G., {Franx}, M., {Kriek}, M., {et~al.} 2008, \apjl, 677, L5

\bibitem[{{van Dokkum} {et~al.}(2010){van Dokkum}, {Whitaker}, {Brammer},
  {Franx}, {Kriek}, {Labb{\'e}}, {Marchesini}, {Quadri}, {Bezanson},
  {Illingworth}, {Muzzin}, {Rudnick}, {Tal}, \& {Wake}}]{vandokkum2010}
{van Dokkum}, P.~G., {Whitaker}, K.~E., {Brammer}, G., {et~al.} 2010, \apj,
  709, 1018

\bibitem[{{Whitaker} {et~al.}(2011){Whitaker}, {Labb{\'e}}, {van Dokkum},
  {Brammer}, {Kriek}, {Marchesini}, {Quadri}, {Franx}, {Muzzin}, {Williams},
  {Bezanson}, {Illingworth}, {Lee}, {Lundgren}, {Nelson}, {Rudnick}, {Tal}, \&
  {Wake}}]{whitaker2011}
{Whitaker}, K.~E., {Labb{\'e}}, I., {van Dokkum}, P.~G., {et~al.} 2011, \apj,
  735, 86

\bibitem[{{Williams} {et~al.}(2009){Williams}, {Quadri}, {Franx}, {van Dokkum},
  \& {Labb{\'e}}}]{williams2009}
{Williams}, R.~J., {Quadri}, R.~F., {Franx}, M., {van Dokkum}, P., \&
  {Labb{\'e}}, I. 2009, \apj, 691, 1879

\bibitem[{{Williams} {et~al.}(2010){Williams}, {Quadri}, {Franx}, {van Dokkum},
  {Toft}, {Kriek}, \& {Labb{\'e}}}]{williams2010}
{Williams}, R.~J., {Quadri}, R.~F., {Franx}, M., {et~al.} 2010, \apj, 713, 738

\bibitem[{{Wuyts} {et~al.}(2011){Wuyts}, {F{\"o}rster Schreiber}, {van der
  Wel}, {Magnelli}, {Guo}, {Genzel}, {Lutz}, {Aussel}, {Barro}, {Berta},
  {Cava}, {Graci{\'a}-Carpio}, {Hathi}, {Huang}, {Kocevski}, {Koekemoer},
  {Lee}, {Le Floc'h}, {McGrath}, {Nordon}, {Popesso}, {Pozzi}, {Riguccini},
  {Rodighiero}, {Saintonge}, \& {Tacconi}}]{wuyts2011}
{Wuyts}, S., {F{\"o}rster Schreiber}, N.~M., {van der Wel}, A., {et~al.} 2011,
  \apj, 742, 96

\bibitem[{{Y{\i}ld{\i}r{\i}m} {et~al.}(2017){Y{\i}ld{\i}r{\i}m}, {van den
  Bosch}, {van de Ven}, {Mart{\'{\i}}n-Navarro}, {Walsh}, {Husemann},
  {G{\"u}ltekin}, \& {Gebhardt}}]{yildirim2017}
{Y{\i}ld{\i}r{\i}m}, A., {van den Bosch}, R.~C.~E., {van de Ven}, G., {et~al.}
  2017, \mnras, 468, 4216

\end{thebibliography}

\end{document}